Signal, image et multimédia

# Compression d'images par SVD et sur-approximation des composantes de chrominance

Henri Bruno Razafindradina, Nicolas Raft Razafindrakoto

L.I.I.S.T.A. (Laboratoire d'Informatique appliquée, d'Images, de Signal, de Télécommunications et d'Automatique)
Ecole Supérieure Polytechnique d'Antananarivo, Université d'Antananarivo, Madagascar
hbazafindradina@gmail.com, raft@nic.mg

**RÉSUMÉ.** Dans cet article, nous présentons un nouveau schéma de compression d'image en couleurs qui effectue une approximation de la matrice des valeurs singulières. L'image doit être convertie dans l'espace luminance / chrominance avant d'être traitée comme dans le cas de la norme JPEG 4 : 2 : 0. Notre algorithme repose sur un sous-échantillonnage de la chrominance, puis une sur-approximation des valeurs singulières de cette dernière. Au lieu de ne retenir que les k premières valeurs singulières pour les 3 composantes R, V et B de l'image, on retient les k premiers coefficients pour la composante Y et seulement k' (k' ≤ k) coefficients pour les 2 composantes $C_b$ et $C_r$. Les résultats ont montré que pour des images de 512×512 pixels, à partir de k = 40 correspondant à une distorsion moyenne de 30 dB et un ratio de 15 : 1, l'image restituée est de bonne qualité. L'algorithme permet aussi un gain de vitesse considérable grâce au sous échantillonnage.

**ABSTRACT.** This paper gives a new scheme of colour image compression related to singular values matrix approximation. The image has to be converted in luminance / chrominance space before being processed like JPEG standard 4 : 2 : 0. Our approach is first based on a chrominance sub-sampling, then an over estimation of its singular values. Instead of keeping only the k first singular values for the 3 components R, G and B, we hold k first coefficients for the Y component and only k' (k' ≤ k) coefficients for 2 components $C_b$ and $C_r$. Results show that for 512×512 pixels that, from k = 40 corresponding in an average distortion of 30 dB and a ratio of 15 : 1, the restored image has good quality. The algorithm allows a significant speed gain by sub-sampling too.

**MOTS-CLÉS :** compression, valeurs singulières, sur-approximation, sous-échantillonnage.

## 1. Introduction

La compression d'images numériques a connu une évolution incessante, parallèlement à celle des télécommunications et du multimédia, depuis les années 60. Elle permet de réduire la taille d'une image dans le but d'augmenter la capacité des supports de stockages (limités en capacité) et d'optimiser l'utilisation de la bande passante d'un réseau. Depuis la normalisation de l'algorithme JPEG basé sur la transformée en cosinus discrète [13], le volume des données multimédias (son, image, vidéo, etc.) n'a cessé d'augmenter. La norme JPEG2000 basée sur la transformée par ondelettes [9][5] a permis d'augmenter le taux de compression des images avec une qualité supérieure à celle de JPEG. L'objet de cet article est la compression d'images en couleur par Décomposition en Valeurs Singulières ou *SVD*. La *SVD* consiste à décomposer une matrice en un produit de 3 matrices *U, S* et *V* (*S* est appelée matrice des valeurs singulières). Chen [7] ainsi qu'Abrahamsen [1] ont déjà proposé une méthode simple de compression d'images à niveaux de gris ne retenant que les *k* premières valeurs singulières. Des améliorations ont été proposées en utilisant l'algorithme *SVD* standard [2]. D'autres applications de la décomposition en valeurs singulières comme la compression faciale [6] ou la reconnaissance faciale [12] ont montré que la *SVD* est utilisée dans plusieurs domaines de l'imagerie. En ce qui concerne les images en couleurs, Adams [3] et Cooper [8] ont proposé une méthode qui applique la compression *SVD* citée plus haut à chaque composante *R*, *V* et *B*.

Dans cet article, nous présentons une nouvelle méthode de compression par décomposition en valeurs singulières qui, avant le codage *SVD* par sur-approximation, sous-échantillonne les composantes chromatiques de l'image. Les bases de la décomposition sont d'abord décrites, nous détaillerons ensuite l'algorithme proposé et les résultats seront discutés.

## 2. La décomposition en valeurs singulières

### 2.1. Principe

Toute matrice *I* de taille $m \times n$ de rang *r* peut être décomposée en une somme pondérée de matrices unitaires $m \times n$ par Décomposition en Valeurs Singulières. Les matrices *U* et *V* sont unitaires et *I* peut donc s'écrire :

$$I = U \times S \times V^T = \sum_{i=1}^{n}\left(\sigma_i \times u_i \times v_i^T\right) \qquad (1)$$

Où $S$ est une matrice dont les $r$ premiers termes diagonaux sont positifs, tous les autres étant nuls. Les $r$ termes $\sigma_i$ non nuls sont appelés valeurs singulières de $I$.

$$S = \begin{pmatrix} \sigma_i & \cdots & 0 \\ \vdots & \ddots & \vdots \\ 0 & \cdots & \sigma_n \end{pmatrix} \quad (2)$$

avec $\sigma_1 \geq \sigma_2 \geq \ldots \geq \sigma_r$ et $\sigma_{r+1} \geq \sigma_{r+2} \geq \ldots \geq \sigma_n = 0$

## 2.2. Application de la SVD en compression d'images

Les valeurs singulières représentent l'énergie [11] de l'image. En effet, l'énergie totale de l'image $I$ est représentée dans la formule suivante :

$$\|I\| = trace[I^T \times I] = \sum_{i=1}^{m}\sum_{j=1}^{n} I^2(i,j) = \sum_{i=1}^{n} \sigma_i^2 \quad (3)$$

La compression d'une image à niveaux de gris vient donc intuitivement en forçant les valeurs singulières les plus faibles à zéro. En ne sélectionnant que les $k$ ($k \leq n$) premières valeurs singulières, on peut approximer la matrice $I$ par la formule :

$$I_k = U \times S_k \times V^T = \sum_{i=1}^{k} \sigma_i \times u_i \times v_i^T \quad (4)$$

$S_k$ représente la matrice des $k$ valeurs singulières compressée. Le produit $U \times S_k$ représente la composante principale [4] de l'image. L'énergie correspondante est :

$$\|I_k\| = trace[I_k^T \times I_k] = \sum_{i=1}^{k} \sigma_i^2 \quad (5)$$

Notons *SVD-k* le codage d'une image à l'aide de $k$ valeurs singulières.

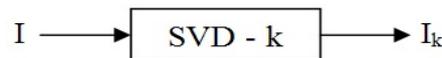

**Figure 1.** Codage SVD pour un nombre k de valeurs singulières

## 2.3. Ratio

Le ratio $G$ du codage est exprimé par la formule suivante :

$$G = (m \times n) / [k \times (m + n + 1)] \quad (6)$$

Comme $G$ doit être supérieur à 1, il faut donc choisir $k$ tel que :

$$k < (m \times n)/(m + n + 1) = k_{seuil}$$

Pour la compression d'images couleurs, Copper [8] a proposé d'appliquer la même approximation aux 3 composantes *R*, *V* et *B*. Dans ce cas, le ratio est égal à *G* de l'équation (6).

## 2.4. Schéma de compression proposé

Sachant que pour une image couleur, l'œil est beaucoup plus sensible aux variations de luminance qu'à celles de couleur, nous proposons une méthode de compression qui agit dans l'espace Luminance / Chrominance ($Y / C_bC_r$ ou $Y / UV$). Au lieu de ne retenir que *k* valeurs singulières pour les 3 plans *RVB* [8], notre méthode traite indifféremment le plan *Y* et les 2 autres plans. Le codage nécessite dans ce cas un prétraitement qui est le passage de *RVB* à $YC_bC_r$. Les 2 plans $C_b$ et $C_r$ sont ensuite sous-échantillonnés suivant le plan horizontal puis vertical.

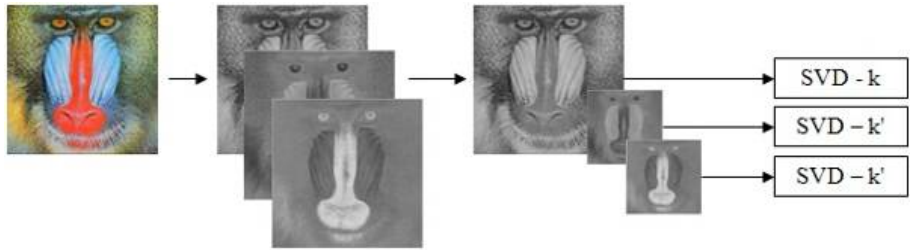

Image originale    Dans l'espace *YCbCr*    Sous-échantillonnage    Codage *SVD*

**Figure 2.** Schéma du codage SVD proposé

La figure 2 représente le schéma de compression proposé : le plan *Y* est codé avec un nombre *k* de valeurs singulières alors que les 2 autres sont codés avec un nombre $k' \leq k$.

Dans ce cas, l'expression du ratio devient :
$$G = (3 \times m \times n)/\{[k \times (m + n + 1)] + [k' \times (m + n + 2)]\} \qquad (7)$$

## 3. Résultats

Tous les tests ont été effectués sur les images « lena, mandrill et plane » de dimension $512 \times 512$ pixels pour plusieurs valeurs de *k*. Et $k' = k / v$, où v est un multiple de *2*.

Le PSNR (Peak Signal to Noise Ratio) a été choisi pour mesurer la distorsion correspondant à un ratio donné. Soit $I$ l'image originale et $I_k$ l'image compressée.

$$PSNR = 10 \times \log_{10}\left(\frac{Max[I(i,j)]^2}{EQM}\right) \quad (8)$$

où EQM est l'erreur quadratique moyenne :

$$EQM = \frac{1}{m \times n}\sum_{i=1}^{m}\sum_{j=1}^{n}[I(i,j) - I_k(i,j)]^2 \quad (9)$$

Les figures suivantes montrent respectivement, pour $k = 20$ et $k = 40$, l'image originale, celle compressée à $k' = k$, puis $k' = k / 2$ et finalement pour $k' = k / 4$.

Pour $k = 20$ :

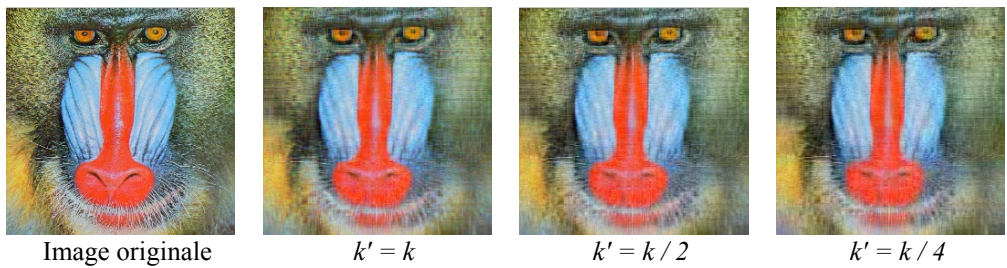

Image originale    $k' = k$    $k' = k / 2$    $k' = k / 4$

**Figure 3.** Résultat obtenu pour k = 20

Pour $k = 40$ :

Image originale    $k' = k$    $k' = k / 2$    $k' = k / 4$

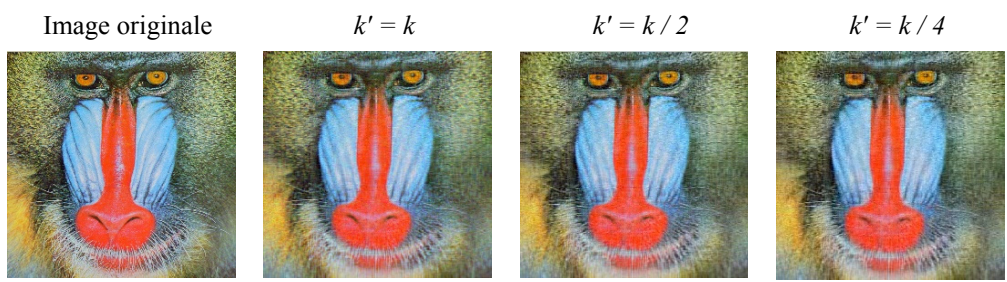

**Figure 4.** Résultat obtenu pour k = 40

Nous avons fait varier ν jusqu'à *16* et les courbes 5 et 6 suivantes présentent respectivement les variations du *PSNR* et du ratio *G* en fonction de *k*.

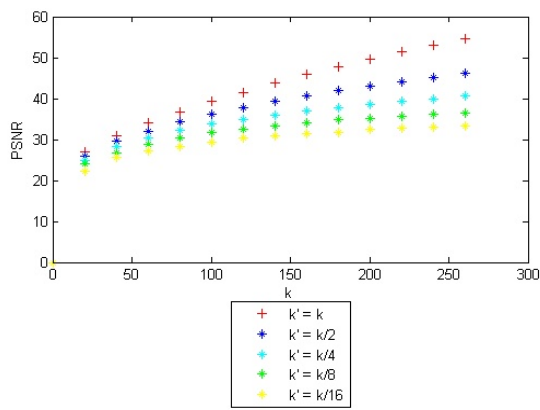

**Figure 5.** Variation du PSNR en fonction de k

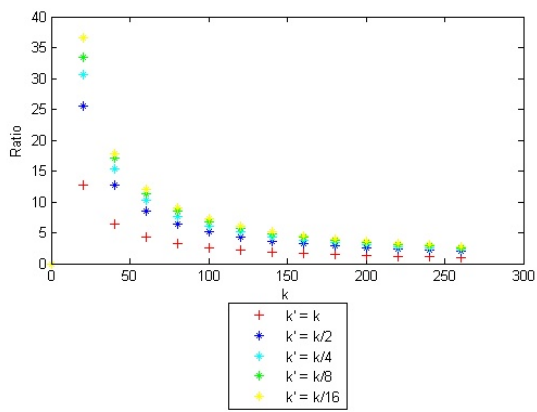

**Figure 6.** Variation du Ratio en fonction de k

Le rapport entre la vitesse de codage dans le système de Cooper et celle de notre approche est résumé dans le tableau suivant :

| Images de test   | Mandrill | Lena | Plane |
|------------------|----------|------|-------|
| Rapport de vitesse | 2.8    | 2.9  | 2.8   |

**Tableau 1.** Rapport de vitesse de codage par rapport au schéma de Cooper

Enfin, nous avons comparé l'énergie de l'image codée avec celle de l'image originale. La courbe suivante montre la variation de ce rapport en fonction de $k$.

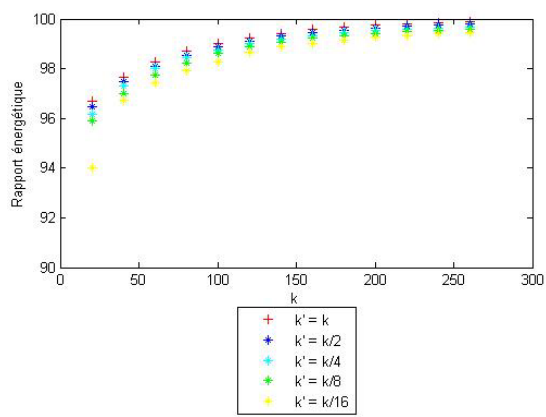

**Figure 7.** Variation du Rapport énergétique (%) en fonction de k

Le cas $k' = k$ correspond à la méthode de Cooper [8].

## 4. Discussion

Les extraits d'images compressées nous montrent qu'à partir de $k = 40$, l'énergie de l'image est presque reconstruite. En effet, sur la courbe de variation du *PSNR*, nous pouvons constater qu'à cette valeur de *k*, la distorsion moyenne obtenue est *30 dB*. La courbe de la figure 7 montre qu'à partir de $k = 40$, l'image codée contient *98 %* de l'énergie de l'image originale pour toutes les valeurs de *k'*.

Le fait de sur-approximer les composantes de chrominance améliore grandement le ratio et la vitesse de compression. En effet, pour $k = 40$, le ratio (figure 6) fluctue autour de *15 : 1* au lieu de *6 : 1* dans le schéma de Cooper et le tableau 1 donne un rapport de vitesse moyen de 3. Plus la taille des composantes de chrominance diminue, plus le gain en vitesse augmente ; la vitesse de codage est donc inversement proportionnelle à *k'*. Vu la capacité de calcul demandée par la SVD, cette vitesse diminue quand on augmente la taille des images à coder. En plus, pour tous les *k* et *k'* la courbe du rapport énergétique reste très proche de celle obtenue avec l'algorithme de Cooper [8].

Malgré ces avantages, la méthode possède des limites : le choix de *k'* est limité car si on dépasse la valeur *16* pour ν, le *PSNR* descendrait à un niveau inférieur à *30*. Le processus de compression est assez lent car le cœur du codage est la décomposition en valeurs singulières qui nécessite un temps de calcul très important mais le décodage est tout aussi rapide que les autres algorithmes tels que JPEG, JPEG2000, etc.

## 5. Conclusion et perspectives

Cet article nous a montré l'efficacité de la décomposition en valeurs singulières en compression d'images couleurs. Avec la puissance de calcul des ordinateurs qui ne cessent d'augmenter, l'approche par *SVD* a encore un bel avenir. Notre méthode pourrait être améliorée en associant la *SVD* avec l'Analyse en Composantes Principales.

## 6. Bibliographie

[1] Abrahamsen A., Richards D., « Image Compression using Singular Value Decomposition », *Linear algebra applications*, 2001, 1-14.

[2] Abriham R. et al, « A Variation on SVD Based Image Compression », *Third Workshop on Computer Vision, Graphics, and Image Processing*, 2006, 1-6.

[3] Adams B., Manual N., « Using the Singular Value Decomposition Particularly for the Compression of Color Images »*, College of the Redwoods*, 2005, 1-20.

[4] Agarwal R., Santhnam M.S., « Digital Watermarking in the Singular Vector Domain », *arXiv:cs/0603130v1*, 2006.

[5] Al Abudi B. K., George L. A., « Color Image Compression Using Wavelet Transform », *GVIP 05 Conference,* Cairo, Egypt, 2005, 1-7.

[6] Bryt O., Elat M., « Compression of Facial Images Using the K-SVD Algorithm », *The IEEE 25-th Convention of Electrical and Electronics Engineers in Israel*, Eilat Israel, 2008, 1-29.

[7] Chen J., (2000), « Image compression with SVD », *ECS 289K Scientific Computation*, 13.

[8] Cooper I., Lorenc C., « Image Compression Using Singular Value Decomposition », *College of the Redwoods*, 2006, 1-22.

[9] Meadows S. C., « Color image compression using wavelet transform », Thesis in Electrical Engineering, 1997, 1-86.

[10] Press W., Teukolsky, Vetterling W., Flannery B., « Numerical recipes : the art of scientific computing », *Cambridge University Press*, 1992, 59-70.

[11] Roue B., Bas P., Le Bihan N., « Décomposition et codage hypercomplexes des images couleurs », Mémoire de DEA laboratoire, Grenoble, Laboratoire LIS, 2003.

[12] Thaahirah S. M. Rasied, Othman O. K., Yuslina B. K., « Human Face Recognition Based on Singular value Decomposition and Neural Network », *GVIP 05 Conference,* Cairo, Egypt, 2005, 1-6.

[13] Watson A. B., « Image Compression Using the Discrete Cosine Transform », *Mathematica Journal*, 1994, 81-88.